\documentclass[lettersize,journal]{IEEEtran}
\usepackage{amsmath,amsfonts}
\usepackage{algorithmic}
\usepackage{algorithm}
\usepackage{array}
\usepackage[caption=false,font=normalsize,labelfont=sf,textfont=sf]{subfig}
\usepackage{textcomp}
\usepackage{stfloats}
\usepackage{url}
\usepackage{verbatim}
\usepackage{graphicx}
\usepackage{cite}
\usepackage[colorlinks = true, hidelinks = true]{hyperref}

\usepackage{cite}
\usepackage{multirow}
\usepackage{booktabs}
\usepackage{amsmath,amssymb,amsfonts}
\usepackage{algorithmic}
\usepackage{graphicx}
\usepackage{textcomp}

\begin{document}
\title{3D Wavelet Latent Diffusion Model for Whole-Body MR-to-CT Modality Translation
}
\author{Jiaxu Zheng$^\dagger$, Meiman He$^\dagger$, Xuhui Tang, Xiong Wang, Tuoyu Cao, Tianyi Zeng,  Lichi Zhang, and Chenyu You
\thanks{$^\dagger$ Equally contributed to this work.}
\thanks{Corresponding authors: Tianyi Zeng and Lichi Zhang.}
\thanks{Jiaxu Zheng, Lichi Zhang are with the School of Biomedical Engineering, Shanghai Jiao Tong University, Shanghai 200240, China (e-mail: jiaxu.zheng@sjtu.edu.cn; lichizhang@sjtu.edu.cn).}
\thanks{Meiman He, Xiong Wang are with the School of Information Science and Technology, ShanghaiTech University, Shanghai 201210, China (e-mail: hemm2022@shanghaitech.edu.cn;  
\mbox{wangxiong@shanghaitech.edu.cn}). }
\thanks{Xuhui Tang is with the School of Biomedical Engineering, ShanghaiTech University, Shanghai 201210, China (e-mail: tangxh2023@shanghaitech.edu.cn).}
\thanks{Tuoyu Cao is with the Shanghai United Imaging Healthcare Co., Ltd., Shanghai 201807, China (e-mail: tuoyu.cao@united-imaging.com).}
\thanks{Tianyi Zeng is with the Department of Radiology and Biomedical Imaging, Yale University, New Haven, CT, United States of America (e-mail: tianyi.zeng@yale.edu). }
\thanks{Chenyu You is with the Department of Applied Mathematics \& Statistics and Department of Computer Science, Stony Brook University, Stony Brook, NY, United States of America (e-mail: chenyu.you@stonybrook.edu). }}

\IEEEpubid{0000--0000/00\$00.00~\copyright~2021 IEEE}

\maketitle
\begin{abstract}
Magnetic Resonance (MR) imaging plays an essential role in contemporary clinical diagnostics. It is increasingly integrated into advanced therapeutic workflows, such as hybrid Positron Emission Tomography/Magnetic Resonance (PET/MR) imaging and MR-only radiation therapy. These integrated approaches are critically dependent on accurate estimation of radiation attenuation, which is typically facilitated by synthesizing Computed Tomography (CT) images from MR scans to generate attenuation maps. However, existing MR-to-CT synthesis methods for whole-body imaging often suffer from poor spatial alignment between the generated CT and input MR images, and insufficient image quality for reliable use in downstream clinical tasks.
In this paper, we present a novel 3D Wavelet Latent Diffusion Model (3D-WLDM) that addresses these limitations by performing modality translation in a learned latent space. By incorporating a Wavelet Residual Module into the encoder-decoder architecture, we enhance the capture and reconstruction of fine-scale features across image and latent spaces. To preserve anatomical integrity during the diffusion process, we disentangle structural and modality-specific characteristics and anchor the structural component to prevent warping. We also introduce a Dual Skip Connection Attention mechanism within the diffusion model, enabling the generation of high-resolution CT images with improved representation of bony structures and soft-tissue contrast.
Quantitative assessments demonstrate that our method, 3D-WLDM, achieves superior results, with PSNR improvements of up to 3.98 dB (1.04 dB over the best baseline), SSIM improvements of up to 0.36 (0.02 over the best baseline), and an MAE reduction of up to 53.76 (7.76 lower than the best baseline). Qualitative evaluations and clinical utility assessments using an open-source organ segmentation model further reveal substantial gains in segmentation accuracy, highlighting the translational potential of our method for radiation planning, hybrid imaging, and broader biomedical applications requiring high-fidelity MR-to-CT synthesis. The source code for this work is publicly available on GitHub at \url{https://github.com/hmm823/3D-Wavelet-Latent-Diffusion-Model}.

\end{abstract}

\begin{IEEEkeywords}
MR-to-CT synthesis, cross-modality translation, medical image generation, diffusion model, feature decouple.
\end{IEEEkeywords}

\section{Introduction}
\IEEEpubidadjcol 			 
\label{sec:introduction}

\IEEEPARstart{M}{agnetic} Resonance Imaging (MRI) provides superior soft tissue contrast compared to Computed Tomography (CT) and does not involve ionizing radiation, making it a preferred modality for tumor characterization, early disease detection, and strategic treatment planning \cite{overcast2021advanced, ranjbarzadeh2023brain}. Leveraging these strengths, MRI has been increasingly integrated into advanced clinical applications, including PET/MR hybrid imaging and MR-only radiation therapy. However, both applications require accurate attenuation correction to compensate for radiation absorption in tissue, where CT remains the clinical gold standard due to its ability to directly provide electron density information. In contrast, MRI lacks such data, thereby presenting a significant limitation for PET attenuation correction and dose calculation in MR-only radiotherapy workflows \cite{catana2020attenuation, owrangi2018mri}.
To address this gap, deep learning-based methods for MRI-to-CT synthesis have gained substantial attention \cite{kang2021synthetic, yang2020unsupervised}. These approaches aim to generate synthetic CT images from MR data, enabling the estimation of attenuation maps suitable for PET reconstruction and radiotherapy planning. Additionally, the resulting synthetic CTs facilitate cross-modal registration by transforming heterogeneous alignment tasks into intra-modality problems, thereby improving the accuracy of longitudinal monitoring and lesion tracking \cite{han2022deformable}.

Despite these advances, most existing MR-to-CT synthesis efforts focus on specific body regions such as the head or pelvis. Whole-body MR-to-CT synthesis remains underexplored, despite its clinical relevance in scenarios like total-body PET/MR imaging. Two major challenges have hindered progress in this domain: The first challenge lies in achieving accurate organ and tissue contrast in synthesized CT images. The nonlinear mapping between MRI and CT, rooted in their fundamentally different imaging physics, makes it difficult to achieve the precise reconstruction of fine-grained anatomical structures, such as bones. Effective translation of modality-specific features frequently requires integration of both local and global contextual information. However, traditional convolutional neural networks (CNNs), which rely on stacked local filters and pooling operations, are inherently limited in receptive field and contextual representation. Prior work has shown that larger spatial context improves synthesis performance \cite{li2023ct}, and recent methods have sought to capture this advantage through frequency-domain enhancements and attention mechanisms \cite{liu2020liver, pan2024synthetic, wang2024freqgan, he2023low}. However, these approaches are predominantly constrained to organ-specific or region-focused settings, and their scalability to full-body synthesis has yet to be demonstrated.

The second challenge concerns the anatomical misalignment between the MR and CT images, which often arises due to differences in patient posture, physiological motion, and variations in scanner bed geometry during separate acquisitions. These misalignments complicate the training of end-to-end modality translation models, which typically rely on the assumption of spatial correspondence. Complex anatomical structures and nonlinear distortions further exacerbate this issue. While CycleGAN-based methods incorporate additional constraints, such as gradients and local descriptors, to address anatomical misalignment \cite{yang2020unsupervised, ge2019unpaired}, they are typically limited to localized anatomical regions and often fail to maintain whole-body anatomical consistency. Furthermore, GAN-based models are vulnerable to mode collapse, which impairs their ability to generalize across diverse anatomical variations, resulting in limited synthesis fidelity and clinical applicability.  

Diffusion-based generative models have recently emerged as a promising alternative for medical image synthesis, demonstrating superior performance over traditional GAN-based methods in terms of stability, fidelity, and diversity \cite{wang2024mutual, meng2024multi, pan2024synthetic, xie2024synthesizing}. The stochasticity of the forward diffusion process inherently reduces the risk of mode collapse, while latent space diffusion provides a natural mechanism for capturing long-range dependencies and contextual features \cite{li2024towards}. These advantages motivate the application of diffusion models to the challenging task of whole-body MR-to-CT synthesis.

In this work, we propose a 3D Wavelet Latent Diffusion Model (3D-WLDM) for high-fidelity whole-body MR-to-CT synthesis. 
The proposed framework leverages the robust image generation capabilities of diffusion models to produce synthetic CT images that surpass existing GAN-based and diffusion-based methods in anatomical fidelity, contrast preservation, and structural consistency.
To address the challenges inherent in this task, we introduce several key innovations. 
First, to enhance the accuracy of organ and tissue contrast in synthesized CT images, we integrate a \textit{Wavelet Residual Module} into the encoder-decoder architecture. 
This module enhances the model’s sensitivity to fine anatomical structures, thereby improving the quality of the encoded latent vectors and the reconstructed images derived from them. 
Second, to ensure structural consistency between input MR and synthetic CT images, we employ \textit{Structure-Modality Disentanglement} in the latent space. 
This technique separates anatomical structures from modality-specific characteristics, preserving structural integrity throughout the iterative diffusion process. 
Additionally, we introduce a \textit{Dual Skip Connection Attention} within the diffusion model to filter redundant information and suppress artifacts arising from MRI-specific textures. 
This mechanism selectively emphasizes structural features critical for accurate CT synthesis while minimizing interference from irrelevant modality details.

Our main contributions are summarized as follows: 
\begin{itemize}
    \item We propose \textbf{3D-WLDM}, a novel diffusion-based framework for whole-body MR-to-CT synthesis, which outperforms existing GAN-based and diffusion-based models in terms of anatomical fidelity and structural consistency.
    \item We develop \textit{Wavelet Residual Module} to improve the quality of encoding and decoding between image and latent spaces, yielding CT images with more precise anatomical structure.     
    \item We introduce \textit{Structure-Modality Disentanglement} to anchor anatomical structure during latent space modalities translation, enhancing spatial consistency under cross-modal diffusion process.
    \item We incorporate \textit{Dual Skip Connection Attention mechanism} to refines the fusion of structural and modality-specific features, resulting in greater synthesis stability and reduced artifact prevalence.
\end{itemize}

\section{Related Works}

\begin{figure*}[htb!]
\centerline{\includegraphics[width=0.95\textwidth]{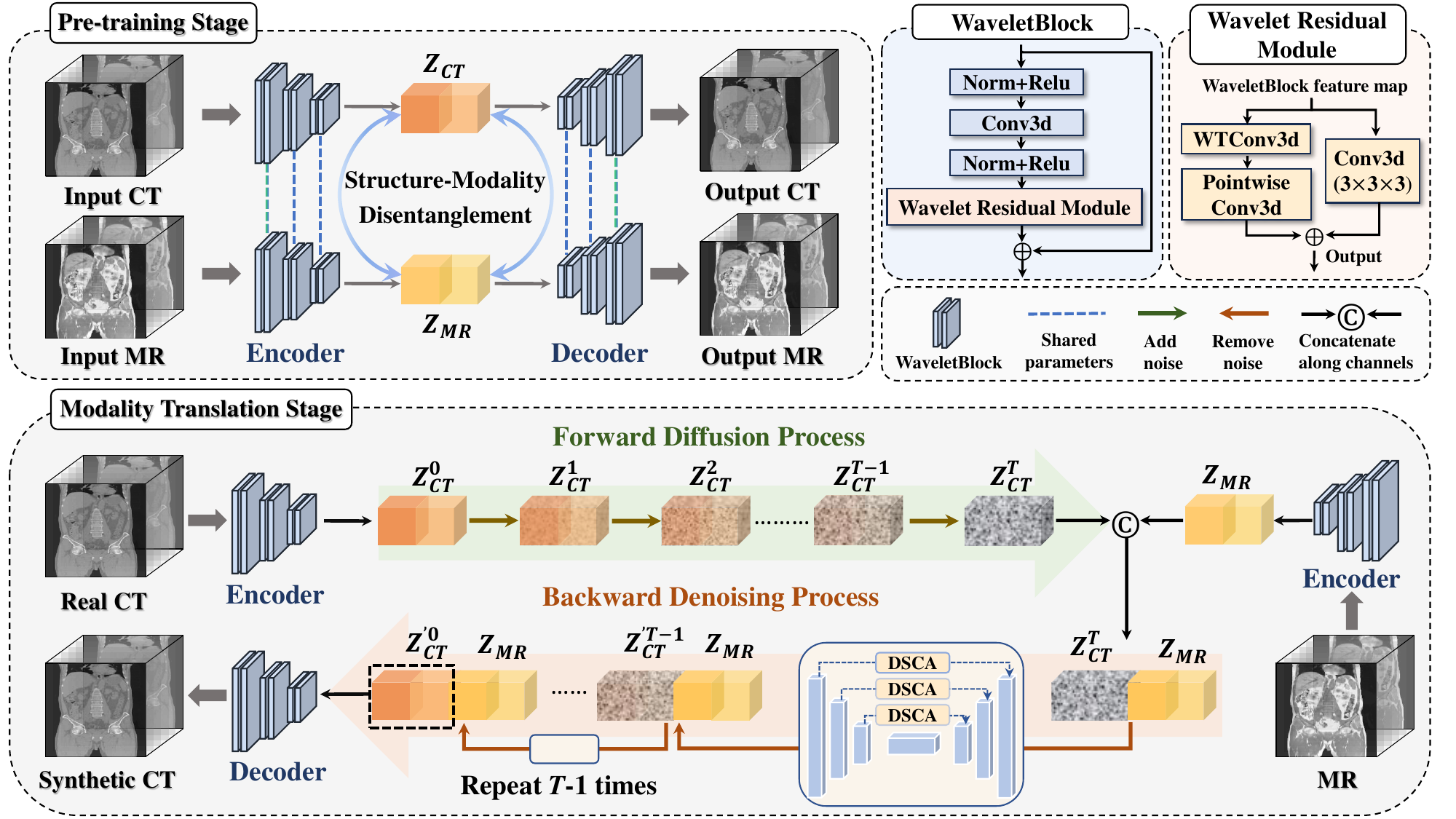}}
\caption{Illustration of the proposed 3D-WLDM architecture. During the pre-training stage, we leverage Structure-Modality Disentanglement to align structural and modality information of the input CT and MR data in the latent space. Additionally, we enhance the response to frequency information by incorporating a Wavelet Residual Module into the WaveletBlocks. During the modality translation stage, the real CT is firstly encoded into the latent feature representation \( Z_{CT}^{0}\), and \( Z_{CT}^{T} \) is subsequently derived through $T$ timesteps of the forward diffusion process. In the backward denoising process, MR is encoded into the latent feature representation \( Z_{M\!R}\), which are concatenated with \( Z_{CT}^{T}\) and fed into a U-net with the Dual Skip Connection Attention (DSCA) to predict the added noise. Through $T$ time steps of iterative denoising, the CT feature representation \( {Z^{\prime}}_{CT}^{0} \) is progressively recovered and decoded into synthetic CT.}
\label{fig1_overview}
\end{figure*}

\subsection{MR-to-CT Modality Translation Methods}

MR-to-CT image synthesis has garnered substantial research attention owing to its critical role in a range of clinical applications, including PET attenuation correction and radiotherapy planning. A variety of modality translation approaches have been proposed, with conditional Generative Adversarial Networks (cGANs) demonstrating particular success in synthesizing CT images from MR data across anatomical regions such as the pelvis, brain, and head \cite{fu2020synthetic, dalmaz2022resvit, florkow2020deep, klaser2021imitation, sun2022synthesis}. However, MR-to-CT translation remains challenging in regions characterized by significant physiological deformation—most notably the abdomen—due to the pronounced spatial misalignment between MR and CT acquisitions \cite{fu2020synthetic, cusumano2020deep}. This issue becomes even more complex in whole-body synthesis tasks, where inter-organ motion and posture variability further exacerbate anatomical inconsistencies.
To mitigate these challenges, CycleGAN-based frameworks have been widely adopted for unpaired cross-modality synthesis, often supplemented with additional spatial constraints such as gradient consistency, mutual information, or anatomical priors \cite{yang2020unsupervised}. For whole-body applications specifically, certain studies have proposed the use of structural contour regularization to preserve global anatomical layout and reduce misalignment artifacts \cite{ge2019unpaired}. Despite these efforts, CycleGAN and its variants remain constrained by inherent limitations of adversarial training—particularly mode collapse, which can reduce synthesis diversity and generalizability. Additionally, the fundamental asymmetry in information content between MR and CT violates the cyclic consistency assumption of CycleGANs, further impairing their efficacy in capturing realistic CT features across large anatomical scales \cite{brou2021improving}.

Cross-modality translation is inherently a high-dimensional, nonlinear, and ill-posed problem \cite{nie2018medical}, which often exceeds the modeling capacity of conventional deep learning frameworks. To enhance synthesis performance, several innovative architectures have been introduced. Shi et al. \cite{shi2021frequency} proposed a frequency decomposition module to separately process high- and low-frequency information, thereby improving anatomical detail preservation. Dovletov et al. \cite{dovletov2022grad, dovletov2023double} introduced a Grad-CAM-guided framework that leverages pre-trained classifiers to focus the generator on diagnostically relevant regions. However, the reliance on convolutional neural networks (CNNs) in these methods limits their ability to capture long-range spatial dependencies and contextual relationships.
To address this, hybrid CNN–Transformer models have been explored. For example, RTCGAN \cite{zhao2023ct} integrates residual Transformers into the generator to extract hierarchical contextual features, while MTT-Net \cite{zhong2023multi} incorporates image token-based Transformers within a 3D generator to model multi-scale spatial correlations. Despite their promising results, these GAN-based methods still face persistent challenges related to spatial misalignment, mode collapse, and limited contextual understanding—highlighting the need for alternative frameworks capable of more robust and anatomically consistent MR-to-CT synthesis.

\subsection{Diffusion-Based Image Synthesis Methods}
Diffusion models have recently demonstrated superior performance over GANs in image-to-image translation tasks by alleviating common issues such as mode collapse \cite{saharia2022palette, chen2024toward}. For instance, StyleDiffusion \cite{wang2023stylediffusion} introduced a framework for content–style disentanglement that explicitly extracts content information while implicitly learning complementary style features. This allows for interpretable and controllable style transfer. To enable flexible conditional generation, ControlNet \cite{zhang2023adding} proposed integrating lightweight adapter modules into pre-trained diffusion models, enabling efficient fine-tuning with external guidance. However, these methods have largely been applied to 2D image synthesis. When naively extended to 3D, they often suffer from inter-slice discontinuities due to the lack of spatial context across adjacent slices.

To overcome this limitation, Pan et al. proposed MC-DDPM \cite{pan2024synthetic}, which integrates shifted-window transformers into a 3D diffusion framework to capture global anatomical structures for MR-to-CT translation. While effective, the model’s sampling process is computationally intensive. In parallel, Friedrich et al. introduced WDM \cite{friedrich2024wdm}, a 3D wavelet-based diffusion model that applies denoising in the wavelet domain. By separating and treating frequency components independently, WDM enhances low-frequency global structures and suppresses high-frequency noise and artifacts. Tapp et al. \cite{tapp2024mr} also explored diffusion-based MR-to-CT synthesis by embedding the denoising process within a 3D CNN architecture; however, the improvements in the synthesis fidelity remains to be modest.

On the other hand, training diffusion models in pixel space remains computationally demanding and leads to slow inference times. To mitigate these issues, Latent Diffusion Models (LDMs) \cite{rombach2022high} were introduced, operating in a lower-dimensional latent space to significantly reduce computational burden. Building on this, Cola-Diff \cite{jiang2023cola} employed cross-attention mechanisms to incorporate diverse conditional signals for multimodal MRI synthesis within the LDMs framework. Despite their efficiency, LDMs methods face challenges in aligning anatomical structures across modalities, particularly for MR-to-CT translation. Misalignment in the latent space and the presence of artifacts in the input MRI can significantly degrade the quality and reliability of the generated CT images.

\section{Methods}
The framework of the proposed 3D Wavelet Latent Diffusion Model (3D-WLDM) is illustrated in Fig.~\ref{fig1_overview}. The 3D-WLDM is designed to generate high-quality synthetic 3D CT images from MR images using a two-stage architecture that integrates a pre-training stage and a modality translation stage. In the pre-training stage, MR and CT images are encoded into latent space using shared-parameter encoders. To enhance the capture and reconstruction of fine-scale features, wavelet blocks with Wavelet Residual Module are incorporated into both the encoders and decoders. Within the latent space, a Structure-Modality Disentanglement mechanism separates structural information from modality-specific details across MR and CT, while aligning the structural information between the two modalities.

Subsequently, the modality translation stage employs a denoising diffusion process in the latent space to facilitate modality conversion from MR to CT, guided by a skip-connection architecture enhanced with Dual Skip Connection Attention (DSCA). This attention mechanism filters out discrepancies in contours and contrast between CT and MR that are propagated through skip connections, reducing interference during the modality conversion process. Finally, the transformed latent representation is decoded back into the image space via the decoder, yielding the synthesized CT image. Together, these components enable 3D-WLDM to effectively handle 3D medical imaging tasks with improved performance.

\subsection{Wavelet Residual Module}
Latent diffusion models typically employ Variational Autoencoders (VAEs) to map images into a compressed latent space, yet this process often compromises structural and contrast details, limiting efficacy in high-fidelity tasks like MR-to-CT modality conversion. To address this, we introduce the Wavelet Residual Module, which leverages wavelet decomposition to separately process high- and low-frequency feature components, thereby enhancing latent feature expressiveness and improving structural fidelity and modality conversion accuracy in the diffusion process.

\begin{figure}[htb]
\centerline{\includegraphics[width=\columnwidth,height=3in]{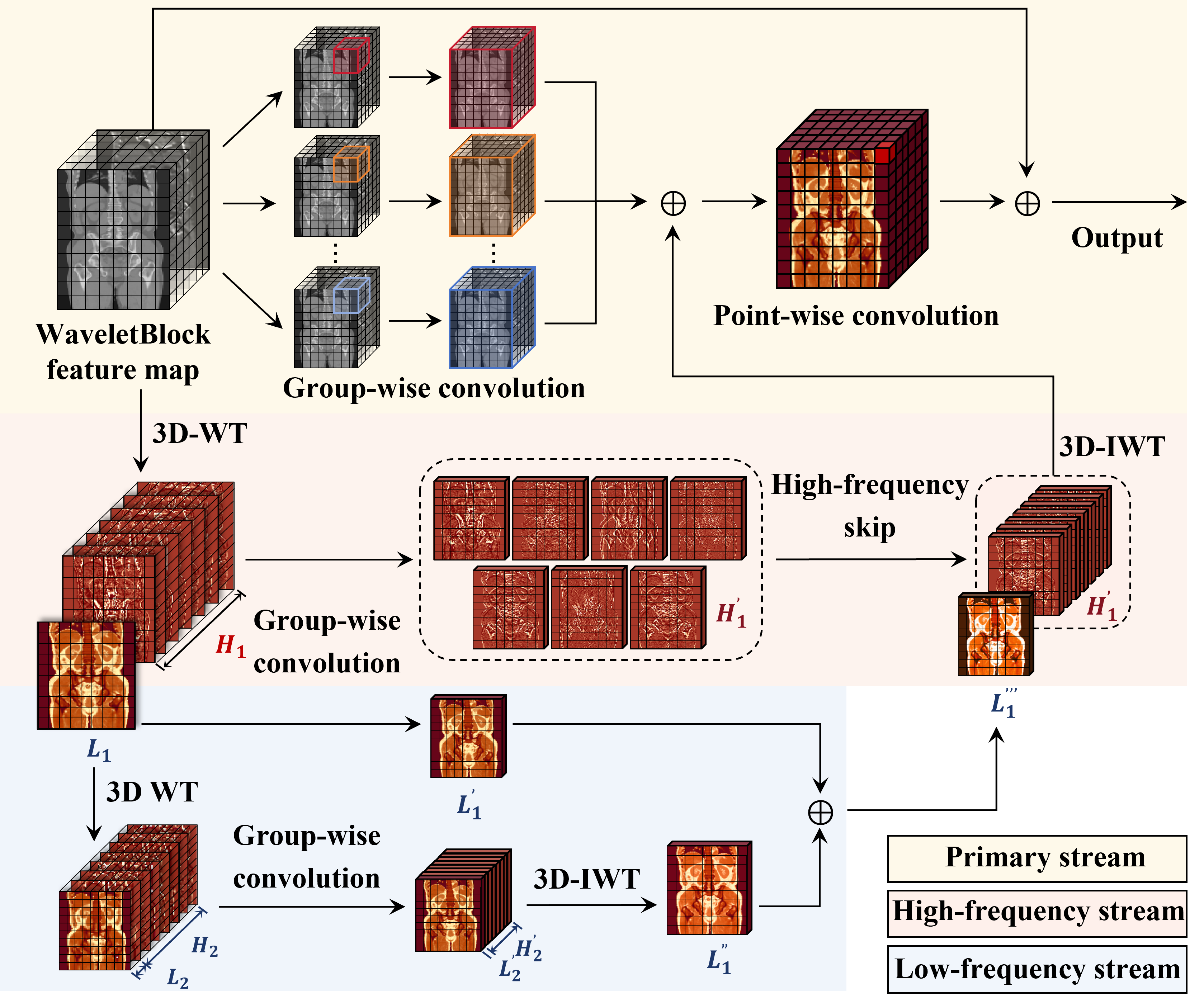}}
\caption{Architecture of the Wavelet Residual Module in the WaveletBlock. The module processes the input feature map through three streams: a primary stream with group-wise and point-wise convolutions, a high-frequency stream generating \( H_1' \), and a low-frequency stream producing \( L_1' + L_1'' \). These two streams are fused via a 3D Inverse Wavelet Transform (3D-IWT) and integrated into the primary stream to enhance the extraction of structural and textural features.}
\label{fig2}
\end{figure}

As shown in Figs.~\ref{fig1_overview} and \ref{fig2}, the Wavelet Residual Module is integrated into the WaveletBlock to enhance feature extraction by incorporating frequency-domain information. The module consists of three interconnected processing streams that manage information from the source domain, low-frequency domain, and high-frequency domain, respectively. The primary stream begins with the input feature map, which is processed through group-wise and point-wise convolutions and augmented by a residual connection to preserve hierarchical features. This stream serves as the backbone for integrating frequency-refined features from the other domains.

In contrast to the primary stream, the second stream is dedicated to high-frequency processing, which starts with the initial 3D Wavelet Transform (3D-WT) that decomposes the input into a high-frequency component \( H_1 \) and a low-frequency component \( L_1 \). The \( H_1 \) component undergoes group-wise convolution to produce \( H_1' \), capturing detailed structural information, which is then held for subsequent fusion. The third stream, focused on low-frequency processing, further decomposes \( L_1 \) using a second 3D-WT into deeper wavelet frequency sub-bands \( L_2 \) (low-frequency) and \( H_2 \) (high-frequency). These sub-bands are processed through group-wise convolutions, and a 3D Inverse Wavelet Transform (3D-IWT) is applied to reconstruct \( L_1'' \), which is then fused with the low-frequency output \( L_1' \) to enhance low-frequency feature extraction.

Once the high-frequency stream (\( H_1' \)) and the low-frequency stream (\( L_1' + L_1'' \)) are obtained, they are subsequently combined and fed into another 3D-IWT to reconstruct the frequency-enhanced feature map. This reconstructed output is then integrated into the primary stream, which undergoes a point-wise convolution to produce the final output. This multi-stream design, leveraging wavelet-based decomposition, group-wise convolutions, and residual learning, increases the capability to encode and decode 3D medical images, contributing to the overall efficacy of the 3D-WLDM framework.

\subsection{Latent Structure-Modality Disentanglement}
Diffusion models rely on long Markov chains, where latent features risk losing structural information over extended iterations. To address this, we propose Structure-Modality Disentanglement, which decouples structural and modality-specific information in latent vectors. This approach anchors structural features during diffusion, enhancing anatomical consistency.

\begin{figure}[h]
    \centering
    \includegraphics[width=\columnwidth]{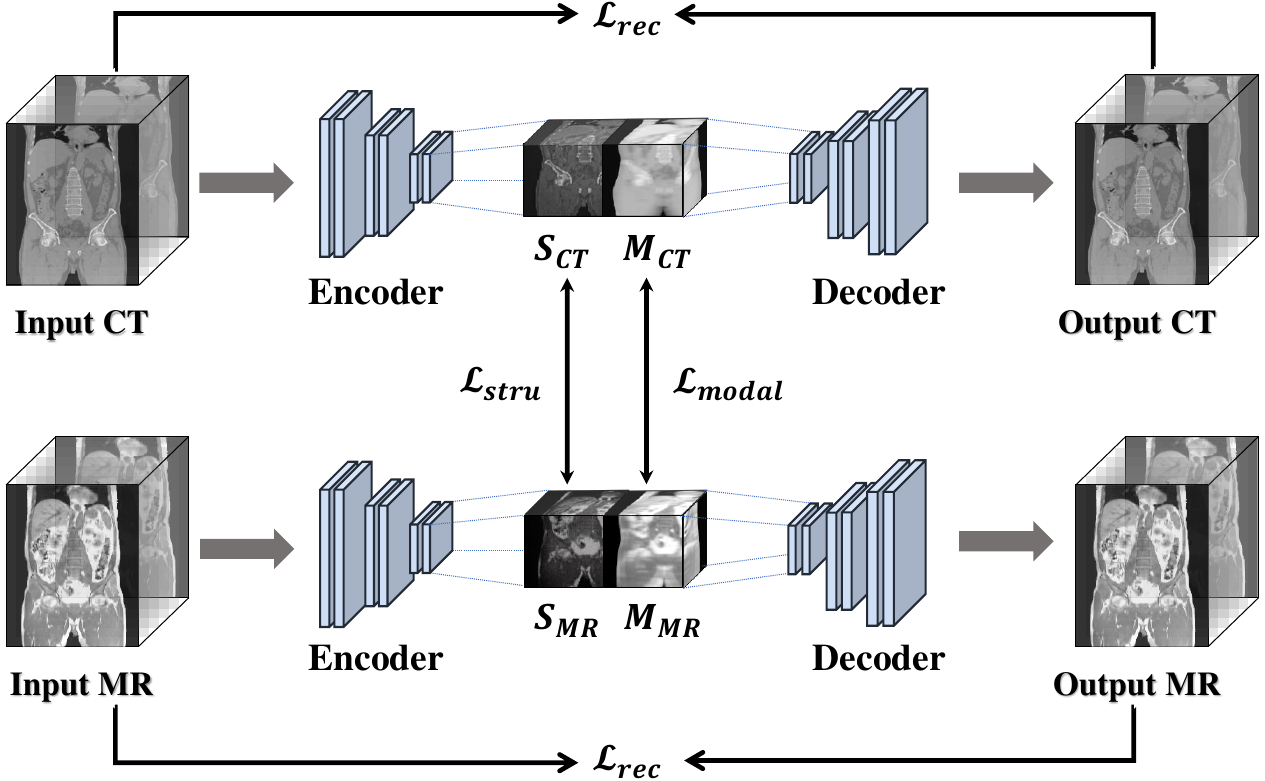}
    \caption{Diagram of the Structure-Modality Disentanglement. The latent vectors \( Z_{CT} \) and \( Z_{M\!R} \) are split into the structural components \( S_{CT} \) and \( S_{M\!R} \) and the modality components \( M_{CT} \) and \( M_{M\!R} \), respectively. The structure loss \( \mathcal{L}_{stru} \) aligns \( S_{CT} \) and \( S_{M\!R} \) for structural similarity, while the modality loss \( \mathcal{L}_{modal} \) enforces dissimilarity between \( M_{CT} \) and \( M_{M\!R} \).}
    \label{fig3}
\end{figure}

The pipeline of Structure-Modality Disentanglement is shown in Fig.~\ref{fig3}, which is proposed to enhance anatomical information in the latent space by decoupling structural information from modality-specific features within latent vectors \( Z \). This decoupling process aims to improve anatomical consistency during the diffusion process, providing a foundation for subsequent information filtering.

The implementation of Latent Structure-Modality Disentanglement is based on the disentanglement loss function, \( \mathcal{L}_D \), which comprises a structure loss \( \mathcal{L}_{stru} \) and a modality loss \( \mathcal{L}_{modal} \). As illustrated in Fig.~\ref{fig3}, we pre-train encoders and decoders using paired MRI and CT volumes from the same patient. For input CT volumes \( I_{CT} \in \mathbb{R}^{D \times H \times W} \) and MRI volumes \( I_{M\!R} \in \mathbb{R}^{D \times H \times W} \), the encoder compresses these inputs along each dimension and outputs their respective distributions. From these distributions, we sample latent feature vectors \( Z_{CT} \in \mathbb{R}^{c \times d \times h \times w} \) for CT and \( Z_{M\!R} \in \mathbb{R}^{c \times d \times h \times w} \) for MRI.

Each latent vector \( Z \) (e.g., \( Z_{CT} \) or \( Z_{M\!R} \)) is explicitly divided into two parts: the structural component \( S \) and the modality component \( M \), such that \( Z = [S, M] \) along the channel dimension. Specifically, we designate the first half of the channels in \( Z_{CT} \) and \( Z_{M\!R} \) to encode their structural attributes as \( S_{CT} \) and \( S_{M\!R} \) respectively, while the second half encodes modality-specific attributes as \( M_{CT} \) and \( M_{M\!R} \). This separation facilitates the disentanglement of anatomical structure (shared across modalities) from modality-specific characteristics (unique to CT or MRI).

To enforce this disentanglement, we define ``paired'' samples as \( Z_{CT} \) and \( Z_{M\!R} \) from the same patient, where the CT and MRI volumes are spatially aligned and thus share structural information. Conversely, ``unpaired'' samples from different patients, such as \( Z_{CT} \) and \( Z_{CT}' \), share the same modality but differ in structure, facilitating the clustering of modality characteristics.

We impose constraints on \( S \) and \( M \) using the disentanglement loss \( \mathcal{L}_D \). The structure loss \( \mathcal{L}_{stru} \) applies a negative cosine similarity constraint within paired samples (e.g., \( S_{CT} \) and \( S_{M\!R} \)) to encourage structural alignment, and a positive cosine similarity constraint between unpaired samples (e.g., \( S_{CT} \) and \( S_{CT}' \)) to minimize structural similarity. Conversely, the modality loss \( \mathcal{L}_{modal} \) enforces dissimilarity within paired samples (e.g., \( M_{CT} \) and \( M_{M\!R} \)) and similarity between unpaired samples (e.g., \( M_{CT} \) and \( M_{CT}' \)) to cluster modality-specific features. These losses are defined as:
\begin{align}
    \mathcal{L}_{stru} &= -\cos \theta_s^{\text{pair}} + \cos \theta_s^{\text{unpair}}, \\
    \mathcal{L}_{modal} &= \cos \theta_m^{\text{pair}} - \cos \theta_m^{\text{unpair}},
\end{align}
where the cosine similarities are:
\begin{align}
    \cos \theta_s^{\text{pair}} &= \frac{S_{CT} \cdot S_{M\!R}}{\|S_{CT}\| \|S_{M\!R}\|} + \frac{S_{CT}' \cdot S_{M\!R}'}{\|S_{CT}'\| \|S_{M\!R}'\|}, \\
    \cos \theta_s^{\text{unpair}} &= \frac{S_{CT} \cdot S_{CT}'}{\|S_{CT}\| \|S_{CT}'\|} + \frac{S_{M\!R} \cdot S_{M\!R}'}{\|S_{M\!R}\| \|S_{M\!R}'\|}, \\
    \cos \theta_m^{\text{pair}} &= \frac{M_{CT} \cdot M_{M\!R}}{\|M_{CT}\| \|M_{M\!R}\|} + \frac{M_{CT}' \cdot M_{M\!R}'}{\|M_{CT}'\| \|M_{M\!R}'\|}, \\
    \cos \theta_m^{\text{unpair}} &= \frac{M_{CT} \cdot M_{CT}'}{\|M_{CT}\| \|M_{CT}'\|} + \frac{M_{M\!R} \cdot M_{M\!R}'}{\|M_{M\!R}\| \|M_{M\!R}'\|}.
\end{align}
The disentanglement loss \( \mathcal{L}_D \) combines these terms as:
\begin{equation}
    \mathcal{L}_D = \mathcal{L}_{stru} + \mathcal{L}_{modal}.
\end{equation}

Here, \( S_{CT} \) and \( S_{M\!R} \) are structural components of paired CT and MRI data, and similarly for \( S_{CT}' \) and \( S_{M\!R}' \). Likewise, \( M_{CT} \) and \( M_{M\!R} \) are modality components of paired data, with \( M_{CT}' \) and \( M_{M\!R}' \) following suit. For unpaired cases, \( S_{CT} \) and \( S_{CT}' \) represent structural components from different CT samples, with analogous definitions for MRI and modality components.

Finally, the encoders and decoders are pre-trained using the total loss function:
\begin{equation}
    \mathcal{L}_\mathit{pre} = \mathcal{L}_{rec} + \alpha \cdot \mathcal{L}_\mathit{KL} + \beta \cdot \mathcal{L}_{D} + \gamma \cdot \mathcal{L}_{adv},
    \label{eq4}
\end{equation}
where \( \mathcal{L}_{rec} \) is the L2 reconstruction loss, \( \mathcal{L}_\mathit{KL} \) aligns the latent distribution with a standard normal prior, and \( \mathcal{L}_{adv} \) is an adversarial loss to enhance generated image quality. The hyperparameters \( \alpha \), \( \beta \), and \( \gamma \) balance the contributions of each term.

\subsection{Dual Skip Connection Attention}
\begin{figure*}[htbp]
\centerline{\includegraphics[width=0.95\textwidth,height=0.25\textwidth]{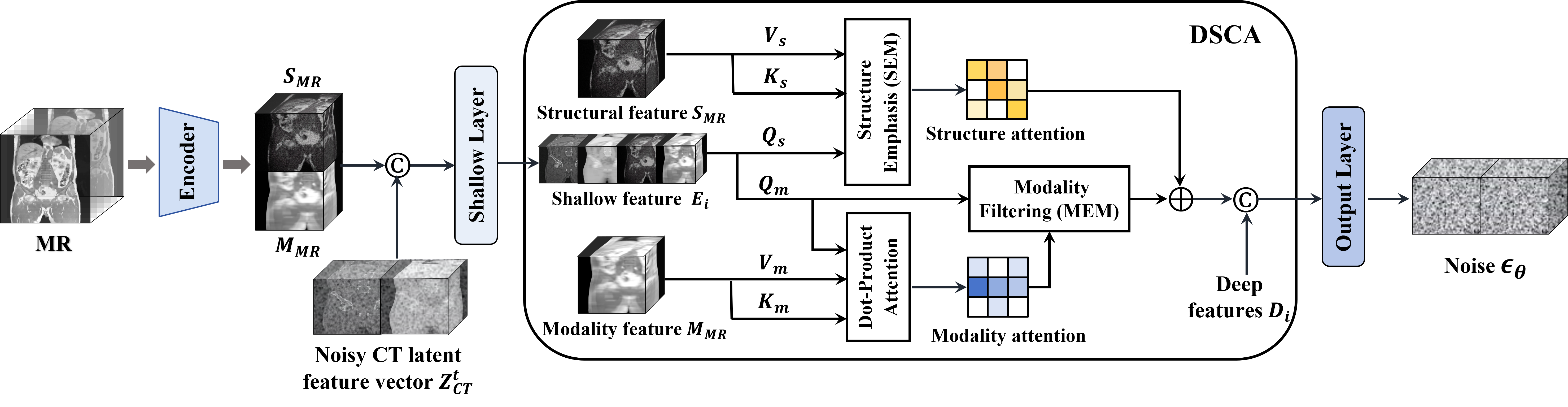}}
\vspace{-5pt}
\caption{Framework of Dual Skip Connection Attention (DSCA) modules in skip connections. Shallow feature map \( E_i \) is derived from the skip connection at the current resolution, while deep feature map \( D_i \) is upsampled from a lower-resolution level.}
\label{fig_unet}
\end{figure*}

MRI offers superior soft tissue contrast compared to CT, but the asymmetric features between these modalities often introduce information redundancy in the CycleGAN frameworks, degrading synthesized CT image quality. To address this, we integrate the Dual Skip Connection Attention module (DSCA) into a skip-connection architecture to filter irrelevant information and enhance structural consistency, as shown in Fig.~\ref{fig_unet}.

The DSCA modules within skip connections at different scales (\(i=1,2,3\)), as depicted in the Fig.~\ref{fig_unet}. Each DSCA module processes three inputs: the MR latent feature \( Z_{M\!R} \), shallow features \( E_i \) from the \(i\)-th encoder layer, and deep features \( D_i \) from the \(i\)-th decoder layer. The DSCA employs a dual cross-attention mechanism, which comprises the Structure Emphasis Module (SEM) and the Modality Filtering Module (MFM) to enhance anatomical accuracy and suppress modality-specific interference, as detailed in the right part of Fig.~\ref{fig_unet}.

The SEM extracts structural information from \( S_{M\!R} \) to improve anatomical accuracy in synthesized CT images. Shallow features \( E_i \) are transformed into a query matrix \( \text{Q}_s \), while \( S_{M\!R} \) generates key matrix \( \text{K}_s \) and value matrix \( \text{V}_s \) matrices. Structural attention \( A_{stru} \) is computed as:
\begin{equation}
A_{stru} = \text{Softmax}\left(\frac{\text{Q}_s \times \text{K}_s^T}{\sqrt{d_{K1}}}\right) \times \text{V}_s,
\label{eq5}
\end{equation}
where \( d_{K1} \) is the dimension of \( \text{K}_s \). The output processed via captures consistent structural features.

The MFM excludes modality-specific interference from \( M_{M\!R} \). Here, \( E_i \) forms the query \( \text{Q}_m \), while \( M_{M\!R} \) produces key \( \text{K}_m \) and value \( \text{V}_m \) matrices. Modality attention \( A_{modal} \) is derived as:
\begin{equation}
A_{modal} = \text{Q}_m - \text{Softmax}\left(\frac{\text{Q}_m \times \text{K}_m^T}{\sqrt{d_{K2}}}\right) \times \text{V}_m,
\label{eq7}
\end{equation}
where \( d_{K2} \) is the dimension of \( \text{K}_m \). The output processed via filters out modality-specific noise.

The SEM and MFM outputs are combined with \( E_i \) via a residual connection, concatenated with \( D_i \), and passed to the next decoder layer. The final DSCA output integrates structurally consistent and modality-filtered features for enhanced synthesis.

\subsection{Condition-guided  Training and Sampling}
The 3D-WLDM framework employs a diffusion model to progressively transform the original CT latent feature \( Z_{CT}^{0} \) into a noisy version, and subsequently recover it through a conditioned reverse process. This process is guided by the MR latent feature \( Z_{M\!R} \) to synthesize high-quality CT images.

In the forward process, Gaussian noise \( \epsilon \) is incrementally added to \( Z_{CT}^{0} \), generating the noisy CT latent feature \( Z_{CT}^{t} \) at time step \( t \), expressed as:
\begin{equation}
Z_{CT}^{t} = \sqrt{\bar{\alpha}_t} \cdot Z_{CT}^{0} + \sqrt{1 - \bar{\alpha}_t} \cdot \epsilon,
\label{eq10}
\end{equation}
where \( \alpha_t = 1 - \beta_t \) and \( \bar{\alpha}_t = \prod_{i=1}^{t} \alpha_i \). The variance \( \beta_t \in (0, 1) \) serves as a hyperparameter controlling the noise level, typically scheduled over \( T \) time steps.

In the conditioned reverse process, the model learns to denoise \( Z_{CT}^{t} \) with guidance from \( Z_{M\!R} \). Specifically, \( Z_{CT}^{t} \) and \( Z_{M\!R} \) are concatenated along the channel dimension and fed into a deeplearning network \( \epsilon_{\theta} \), which is trained using the loss function:
\begin{equation}
\mathcal{L}_\mathit{WLDM} = \mathbb{E}_{\epsilon_{t} \sim \mathcal{N}(0, I)} \left[ \left\| \epsilon_{t} - \epsilon_{\theta} \left( Z_{CT}^{t}, t, Z_{M\!R} \right) \right\|_2^2 \right],
\label{eq11}
\end{equation}
where \( \epsilon_{t} \) is the true noise added at time step \( t \), sampled from a standard normal distribution \( \mathcal{N}(0, I) \), and \( \epsilon_{\theta} \left( Z_{CT}^{t}, t, Z_{M\!R} \right) \) is the noise predicted by the network, conditioned on \( Z_{CT}^{t} \), \( t \), and \( Z_{M\!R} \).

During the sampling phase, the trained network \( \epsilon_{\theta} \) is used to predict the noise at each step, enabling the iterative recovery of \( Z_{CT}^{0} \). For inputs \( Z_{CT}^{t} \) and \( Z_{M\!R} \), the latent feature vector \( Z_{CT}^{t-1} \) for the previous step is computed as:
\begin{equation}
Z_{CT}^{t-1} = \sqrt{\bar{\alpha}_{t-1}} \cdot Z_{CT}' + \sqrt{1 - \bar{\alpha}_{t-1}} \cdot \epsilon_{\theta} \left( Z_{CT}^{t}, t, Z_{M\!R} \right),
\label{eq12}
\end{equation}
where \( Z_{CT}' \) is the intermediate CT latent feature estimated in a single step, derived as:
\begin{equation}
Z_{CT}' = \frac{Z_{CT}^{t} - \sqrt{1 - \bar{\alpha}_t} \cdot \epsilon_{\theta} \left( Z_{CT}^{t}, t, Z_{M\!R} \right)}{\sqrt{\bar{\alpha}_t}}.
\label{eq13}
\end{equation}
This iterative denoising process gradually removes noise, guided by \( Z_{M\!R} \), to reconstruct the clean CT latent feature.

We adopt a fast sampling strategy\cite{song2020denoising} that enables the selection of a subset of time steps for denoising, rather than processing all \( T \) steps sequentially. The recovered denoised CT latent feature \( Z_{CT}^{0} \) is then passed to the decoder to generate the target CT modality data.

\begin{figure*}[htpb]
\centerline{\includegraphics[width=\textwidth,height=8.5in,keepaspectratio]{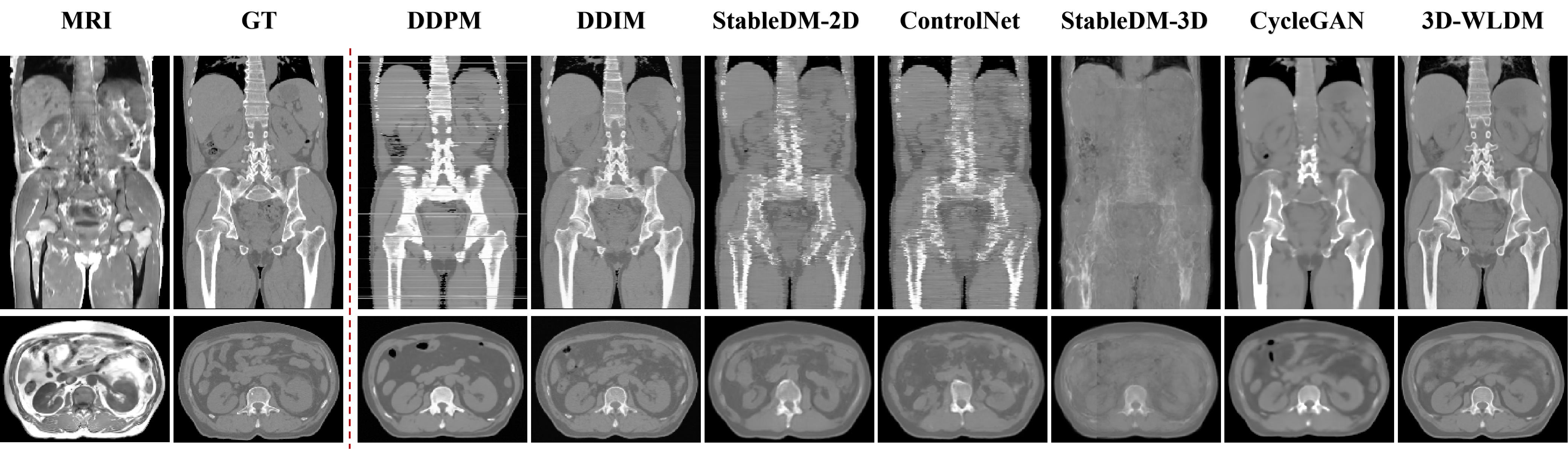}}
\vspace{-5pt}
\caption{Visualization of synthesized CT images in the coronal view (Row 1) and axial view (Row 2). The first column displays the input MRI images, the second column shows the Ground Truth (real CT images), and columns 3 to 9 present the synthesized CT results from various models.}
\label{fig6}
\end{figure*}

\section{Experiments}
\subsection{Dataset and Implementation}
This study was conducted with Institutional Review Board (IRB) approval and utilized a multi-centre dataset comprising 268 subjects recruited from collaborating hospitals. Each subject underwent whole-body PET/MR scanning using a UIH uPMR790 scanner to acquire magnetic resonance (MR) volumes based on the Water-Fat Imaging In-Phase (WFI-IP) sequence, and whole-body PET/CT scanning using a UIH uMI780 scanner to obtain corresponding computed tomography (CT) volumes. The native MR volumes had dimensions of \( 549 \times 384 \times 472 \) with voxel spacing of \( 0.97 \, \text{mm} \times 0.97 \, \text{mm} \times 2.40 \, \text{mm} \); CT volumes were acquired at \( 512 \times 512 \times 900 \) with voxel spacing of \( 0.98 \, \text{mm} \times 0.98 \, \text{mm} \times 1.00 \, \text{mm} \).

To mitigate variability due to upper body positioning, anatomical regions above the shoulders were cropped. Black borders were removed using a rectangular envelope cropping algorithm, and all volumes were resampled to an isotropic voxel spacing of 2$\text{mm}^3$. Intensity normalization was performed for both modalities to range \([-1, 1]\), with CT images truncated to range \([-800, 800]\) HU to enhance contrast in soft tissues. Rigid and non-rigid intermodal registration was performed using Elastix~\cite{klein2009elastix}, employing mutual information as the similarity metric. The final dataset was partitioned randomly into training and testing subsets in a 4:1 ratio.

During training, each volume was randomly cropped into subvolumes of size \( 128 \times 128 \times 128 \). Data augmentation strategies included random flips, brightness mapping, and affine transformations, followed by normalization to \([-1, 1]\). Model training was performed using PyTorch with the Adam optimizer on an NVIDIA A40 GPU. The diffusion process comprised \( T = 1000 \) timesteps, with noise variance linearly scheduled from \( \beta_1 = 0.0001 \) to \( \beta_T = 0.2 \).

\subsection{Comparisons of Multiple Baselines and Discussion}

\subsubsection{Qualitative Analysis} 
We performed MR-to-CT modality translation experiments to benchmark the proposed 3D Wavelet Latent Diffusion Model (3D-WLDM) against several state-of-the-art baselines, including DDPM~\cite{ho2020denoising}, DDIM~\cite{song2020denoising}, StableDiffusion\! 2D (StableDM-2D)\cite{rombach2022high}, ControlNet\cite{zhang2023adding}, StableDiffusion\! 3D (StableDM-3D)\cite{rombach2022high}, and CycleGAN\cite{zhu2017unpaired}. Representative qualitative comparisons are illustrated in Fig.~\ref{fig6}.

In 2D transverse slice generation, DDPM occasionally produces failed outputs, notably horizontal line artifacts in coronal views. Our experiments suggest that DDPM is highly sensitive to the initial noise vector, resulting in substantial variability in output quality. These artifacts are attributed to the model’s limited generative capacity. Increasing the number of diffusion steps reduces such interslice inconsistencies but further exacerbates inference inefficiency, underscoring the trade-off between quality and computational cost in DDPM. DDIM improves upon DDPM by enhancing sampling efficiency and reducing interlayer artifacts. Nevertheless, noise artifacts persist, particularly in transverse-plane reconstructions, suggesting suboptimal structural fidelity.

StableDM-2D and ControlNet generate CT images with visually realistic textures and intensity distributions that closely resemble ground truth. However, both suffer from anatomical misalignment. Structural discontinuities are evident, particularly in axial and coronal planes. These inconsistencies stem from limitations in the variational autoencoder (VAE) decoder, which, while enhancing perceptual quality, degrades structural integrity during latent encoding. The cross-attention mechanism in StableDM-2D and the condition fusion strategy in ControlNet both fall short in enforcing spatial coherence. StableDM-3D amplifies these limitations, yielding overly smooth and anatomically imprecise reconstructions, highlighting the current challenges faced by 3D diffusion frameworks in preserving high-fidelity structural information.

CycleGAN exhibits pronounced anatomical inconsistencies, especially in spinal structures, likely due to ambiguous MR signal characteristics. As a bidirectional generative framework, it tends to propagate non-physiological features introduced by low-contrast MR inputs. These artifacts degrade the fidelity of CT synthesis and compromise structural accuracy.

In contrast, our proposed 3D-WLDM demonstrates marked improvements in both visual fidelity and anatomical coherence. The integration of Wavelet Blocks enhances multiscale encoding and reconstruction, while the Structure-Modality Disentanglement module mitigates inconsistencies in latent space representation. Furthermore, the DSCA module significantly improves the generation of fine structural details during the diffusion process. Together, these components enable 3D-WLDM to produce CT reconstructions that are not only visually compelling but also structurally faithful to the underlying anatomy.

\begin{figure*}[htbp]
\centerline{\includegraphics[width=\textwidth,height=8.5in,keepaspectratio]{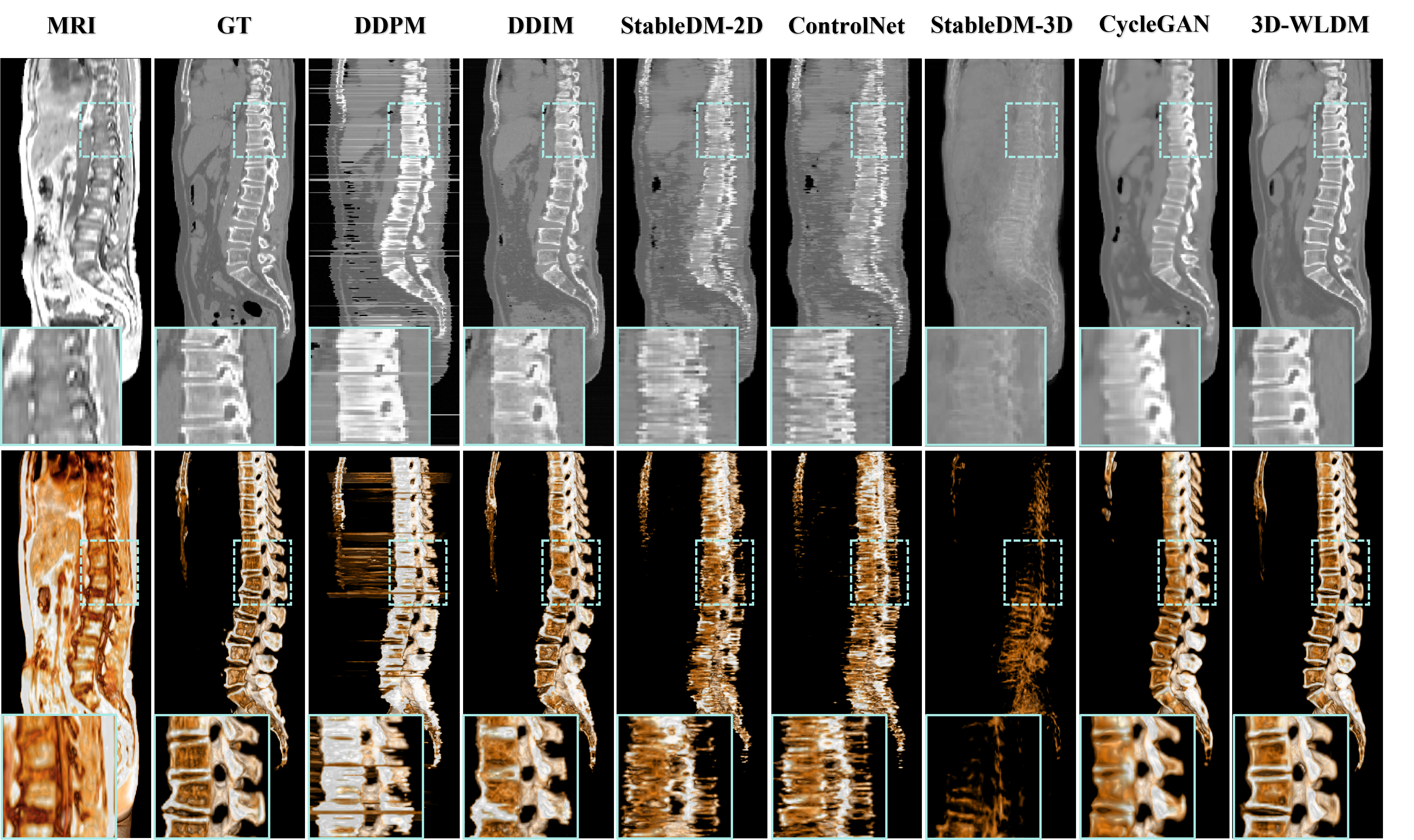}}
\vspace{-5pt}
\caption{Visualization of synthesized CT results on spine. The first row shows slices of the spine in the sagittal view, the second row presents the rendered 3D spine from the generated CT volumes.}
\label{fig_spine}
\end{figure*}

\begin{table}[t]
\centering
\caption{Quantitative Comparisons of Different Methods. Bold values indicate the best performance. ($\downarrow$ denotes lower is better; $\uparrow$\!\!\!\! denotes higher is better).}
\label{tab:comparison}
\renewcommand{\arraystretch}{1.2}
\begin{tabular*}{\columnwidth}{@{\extracolsep{\fill}}l|cccc@{}}
\specialrule{0.8pt}{0pt}{0pt}
\multicolumn{1}{c|}{\multirow{2}{*}{\begin{tabular}[c]{@{}c@{}} Method \end{tabular}}} & 
\multicolumn{4}{c}{Performance metrics} \\ \cline{2-5} 
\multicolumn{1}{c|}{} & 
PSNR(↑) & SSIM(↑) & MAE (↓) & NCC(↑) \\
\specialrule{0.8pt}{0pt}{0pt}
DDPM & 21.35$\pm$3.75 & 0.60$\pm$0.10 & 101.04$\pm$45.52 & 0.93$\pm$0.11 \\
DDIM & 24.29$\pm$2.34 & 0.45$\pm$0.07 & 76.64$\pm$16.00 & 0.97$\pm$0.02 \\
StableDM-2D & 22.59$\pm$2.29 & 0.71$\pm$0.07 & 67.52$\pm$18.64 & 0.95$\pm$0.02 \\
ControlNet & 22.64$\pm$2.18 & 0.73$\pm$0.07 & 66.72$\pm$18.24 & 0.95$\pm$0.02 \\
StableDM-3D & 21.82$\pm$1.74 & 0.66$\pm$0.07 & 85.44$\pm$21.92 & 0.95$\pm$0.02 \\
CycleGAN & 23.44$\pm$2.31 & 0.79$\pm$0.06 & 55.04$\pm$17.28 & 0.96$\pm$0.02 \\
3D-WLDM & \textbf{25.33$\pm$2.60} & \textbf{0.81$\pm$0.07} & \textbf{47.28$\pm$16.48} & \textbf{0.97$\pm$0.02} \\
\specialrule{0.8pt}{0pt}{0pt}
\end{tabular*}
\end{table}

\subsubsection{Quantitative Analysis} 
Table~\ref{tab:comparison} summarizes the quantitative performance of all methods using four standard metrics: Peak Signal-to-Noise Ratio (PSNR), Structural Similarity Index Measure (SSIM), Mean Absolute Error (MAE), and Normalized Cross Correlation (NCC). The proposed 3D-WLDM achieves consistent improvements across all metrics, with PSNR gains of up to 3.98 dB (1.04 dB above the next best), SSIM improvements of 0.36 (0.02 above the next best), MAE reductions of 53.76 (7.76 below the next best), and NCC increases of 0.04, highlighting its superiority in generating anatomically accurate and perceptually faithful CT images.

Quantitative results also reveal important discrepancies between visual quality and structural fidelity across baseline methods. For instance, DDIM generates visually sharper images in the coronal plane (Fig.~\ref{fig6}) compared to StableDM-2D and ControlNet, yet achieves lower SSIM and higher MAE. This contrast is attributed to differences in generative pathways: latent vector–based decoding via variational autoencoders (VAEs) typically enhances perceptual quality and contrast but may fail to preserve accurate anatomical detail when compared to pixel-space diffusion outputs.

CycleGAN achieves the highest SSIM and lowest MAE among baseline methods, despite clear structural deficiencies, such as vertebral blurring and discontinuities (Fig.~\ref{fig6}). This performance is primarily due to its strong contrast preservation and capacity for global structural consistency. However, its failure to reconstruct fine bone structures limits its clinical utility -- particularly in applications such as PET/MR attenuation correction, where accurate bone delineation is critical due to its high electron density and poor visibility in MR imaging.

In contrast, 3D-WLDM delivers state-of-the-art performance across all quantitative and qualitative assessments. Its combination of multiscale wavelet encoding, structure-modality disentanglement, and enhanced structural attention enables improved anatomical fidelity, contrast consistency, and noise suppression. These properties underscore the model's potential for clinical translation, particularly in scenarios demanding high-resolution, structure-preserving CT synthesis.

\begin{figure*}[t]
\centerline{\includegraphics[width=\textwidth,height=8.5 in,keepaspectratio]{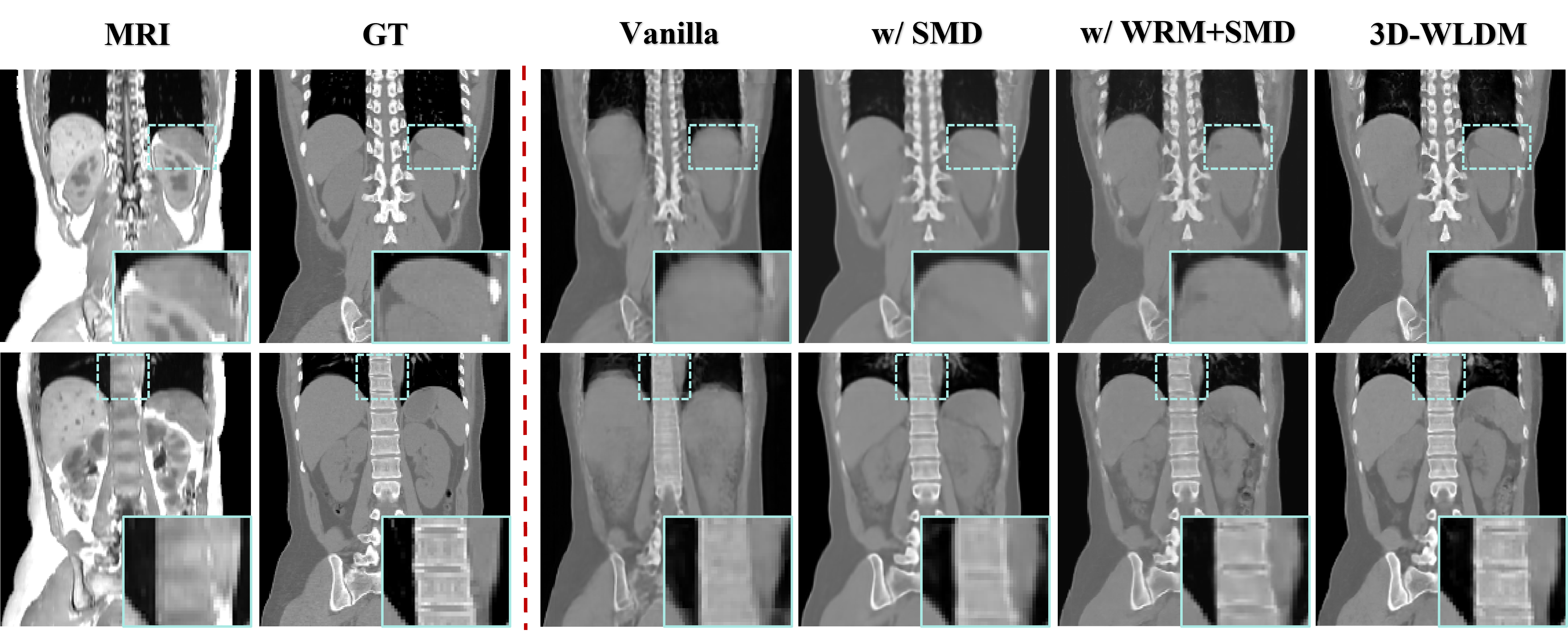}}
\vspace{-5pt}
\caption{Comparison of the results from ablation experiments: Vanilla (3D latent diffusion model), Wavelet Residual Module (WRM), Structure-Modality Disentanglement (SMD), and their combination with DSCA (3D-WLDM).}
\vspace{-5pt}
\label{fig9}
\end{figure*}

\subsubsection{Spine Reconstruction Analysis} 
The spine presents a significant challenge for MR-to-CT synthesis due to the inherently low soft-tissue contrast and blurred cortical boundaries in MR images. These limitations substantially affect the accuracy of vertebral bone reconstruction—an essential component for downstream clinical applications such as radiotherapy planning and PET/MR attenuation correction. To evaluate the fidelity of synthetic vertebral structures, sagittal-plane slices and rendered 3D spine visualizations are shown in Fig.~\ref{fig_spine}, revealing notable differences among competing methods.

Consistent with observations from the axial-plane results in Fig.~\ref{fig6}, both DDPM and DDIM fail to reconstruct anatomically accurate spinal structures. DDPM exhibits poor trabecular bone detail and fails to preserve internal vertebral architecture. Although DDIM shows improved contour alignment, it introduces contrast inversions at intervertebral boundaries, indicating erroneous intensity modeling. These findings suggest that pixel-space diffusion models offer better geometric alignment but are limited in capturing realistic image contrast and bone microstructure.
StableDM-2D and ControlNet produce visually appealing outputs in the axial view, with enhanced contrast and texture fidelity. However, they fail to maintain structural continuity in the spine across sagittal slices, resulting in fragmented and anatomically implausible vertebral columns. This misalignment likely stems from inadequate structural encoding during latent-space diffusion, where spatial correlations between MR inputs and CT targets are poorly preserved.
CycleGAN yields high-contrast outputs but fails to reconstruct the fine trabecular architecture of the spine. This is attributed to the network’s difficulty in modeling cortical bone features, which appear hyperintense and sharply contoured in CT but indistinct in MR. Consequently, large portions of the spine are absent or inaccurately depicted, particularly in coronal slices (Fig.~\ref{fig6}), reducing the method’s utility in bone-critical applications.

In contrast, the proposed 3D-WLDM model achieves markedly improved vertebral bone reconstruction. It preserves global structural alignment with MR inputs while capturing fine-grained details of both cortical and trabecular bone. These results underscore the effectiveness of the model’s wavelet-based multiscale representation and structure-modality disentanglement in overcoming the intrinsic limitations of MR imaging for bone synthesis.

\subsection{Ablations}
\subsubsection{Synthesized Image Analysis} 
To examine the contribution of each model component within 3D-WLDM framework, we performed a series of ablation experiments, as detailed in Table~\ref{tab3}. The Vanilla model serves as the baseline, employing a standard 3D latent diffusion framework in which the MR latent representation is concatenated with Gaussian noise to form a dual-channel input, without any explicit structural or modality-specific enhancements.

Introducing SMD into the baseline enhances the model’s ability to preserve anatomical structure during the generative process. By decoupling structural information from modality-specific features during pretraining, the model learns to anchor geometry consistently across the diffusion trajectory. While this improves overall spatial alignment, some fine-grained details remain suboptimal.

Incorporating WRM further improves image reconstruction by enhancing the representation of high-frequency features. Through wavelet-domain processing, WRM enriches the encoding and decoding between the image and latent spaces, yielding more precise recovery of complex anatomical structures, particularly in osseous regions.

The full model, combining WRM, SMD, and DSCA achieves the most robust performance across all evaluated metrics. DSCA reinforces the interaction between structural and modality-specific signals in the latent space, enabling the network to generate synthetic CT volumes with enhanced stability contrast, reduced artifact. Relative to the baseline, the full model improves PSNR by 2.31 dB, SSIM by 0.23, and reduces MAE by 28.08, underscoring the cumulative benefit of the proposed modules.

\begin{figure*}[t]
\centerline{\includegraphics[width=\textwidth,height=8.5 in,keepaspectratio]{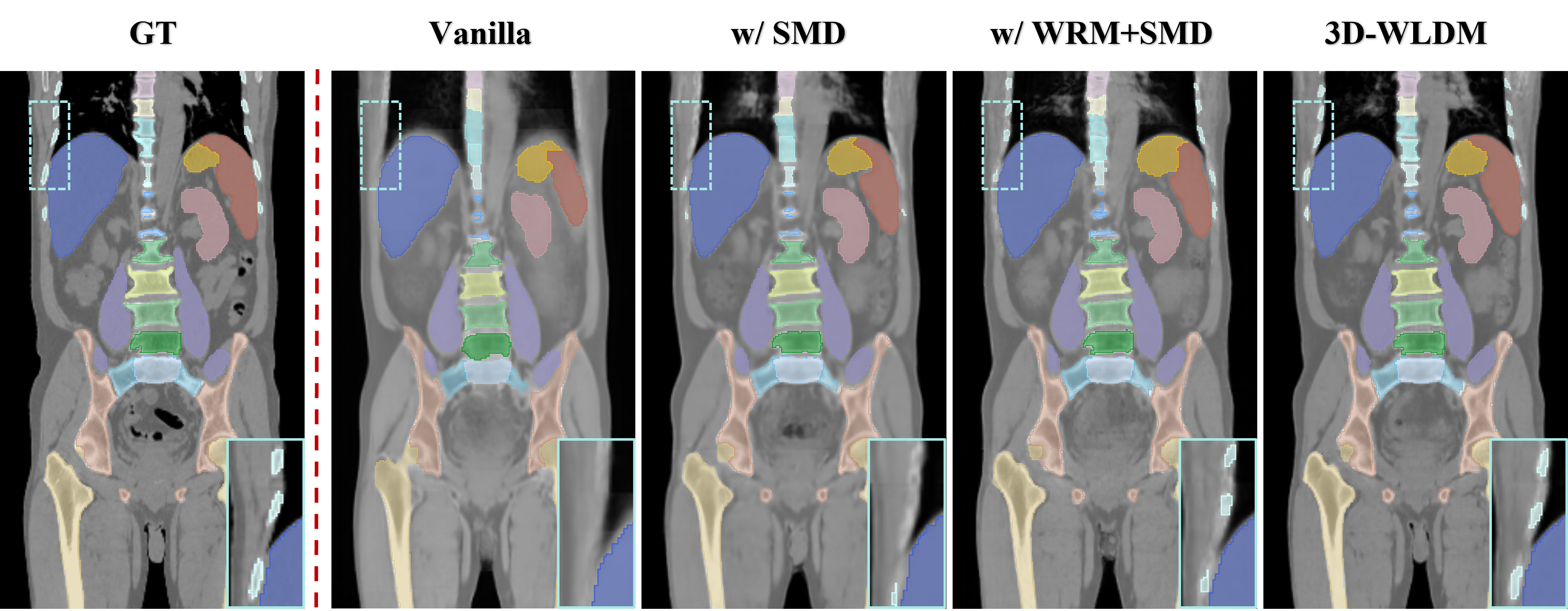}}
\vspace{-5pt}
\caption{The segmentation results of important anatomical structures. GT represents the segmentation results of the real CT, while others represent the segmentation results of the synthesized CT in the ablation experiments.}
\label{fig_sig}
\end{figure*}

\begin{table}[t]
\caption{Ablation on performance of WRM, SMD, and their combination with DSCA (3D-WLDM). Bold values indicate the best performance across configurations. ($\downarrow$ denotes lower is better; $\uparrow$ denotes higher is better).}
\label{tab3}
\centering
\renewcommand{\arraystretch}{1.2}
\setlength{\tabcolsep}{2.0mm}
\begin{tabular*}{\columnwidth}{@{\extracolsep{\fill}}l|cccc@{}}
\specialrule{0.8pt}{0pt}{0pt}
\multicolumn{1}{c|}{\multirow{2}{*}{\begin{tabular}[c]{@{}c@{}} Method \end{tabular}}} & 
\multicolumn{4}{c}{Performance metrics} \\ \cline{2-5} 
\multicolumn{1}{c|}{} & 
PSNR($\uparrow$) & SSIM($\uparrow$) & MAE($\downarrow$) & NCC($\uparrow$) \\ \hline
Vanilla & 23.02$\pm$2.45 & 0.58$\pm$0.09 & 75.36$\pm$23.84 & 0.96$\pm$0.03 \\ 
w/ WRM & 24.65$\pm$2.42 & 0.68$\pm$0.09 & 59.60$\pm$18.80 & 0.97$\pm$0.02 \\ 
w/ WRM+SMD & 25.02$\pm$2.49 & 0.78$\pm$0.10 & 52.24$\pm$19.28 & 0.97$\pm$0.02 \\ 
3D-WLDM & \textbf{25.33$\pm$2.60} & \textbf{0.81$\pm$0.07} & \textbf{47.28$\pm$16.48} & \textbf{0.98$\pm$0.02} \\ 
\specialrule{0.8pt}{0pt}{0pt}
\end{tabular*}
\end{table}

These quantitative findings are further supported by the qualitative results in Fig.\ref{fig9}. The Vanilla model generates synthetic CT images with perceptual quality comparable to StableDM-3D (Fig.\ref{fig6}), yet fails to capture modality-specific transformations with sufficient structural fidelity, particularly in anatomically complex 3D regions where spatial encoding is inherently more challenging.

Introducing SMD leads to more accurate anatomical alignment by explicitly preserving structure throughout the diffusion process. Organ and tissue boundaries become more coherent, though finer anatomical details remain underrepresented. The addition of WRM markedly improves the synthesis of high-frequency features, especially in regions with intricate bone morphology such as the vertebrae and ribs, by leveraging wavelet-domain encoding to reinforce multiscale structural representation.

The full model further benefits from the integration of DSCA, which refines the fusion of structural and modality cues and enhances both generation stability and contrast accuracy. As shown in Fig.~\ref{fig9}, this results in improved delineation of abdominal organs, such as the stomach and kidneys, with sharper boundaries and more realistic texture.

Collectively, these findings substantiate our central hypothesis: that explicit disentanglement of structure and modality, combined with multiscale encoding and attention-guided feature fusion, is critical for achieving accurate, high-quality 3D medical image synthesis in cross-modality tasks.

\begin{table}[h]
\centering
\caption{Ablation on segmentation performance of key structures using WRM, SMD, and their combination with DSCA  (3D-WLDM). Bold values indicate the best performance across configurations. ($\uparrow$\!\!\!\! denotes higher is better).}
\label{tab_seg}
\renewcommand{\arraystretch}{1.3}
\setlength{\tabcolsep}{0.8mm}
\begin{tabular*}{\columnwidth}{@{\extracolsep{\fill}}l|ccccc@{}}
\specialrule{0.8pt}{0pt}{0pt}
\multicolumn{1}{c|}{\multirow{2}{*}{\begin{tabular}[c]{@{}c@{}} Method \end{tabular}}} & 
\multicolumn{5}{c}{DSC($\%$)($\uparrow$)} \\ \cline{2-6} 
\multicolumn{1}{c|}{}
& Spleen             & Kidney             & Liver              & Stomach            & Vertebrae          \\ \hline
Vanilla                        & 64.2$\pm$17.9          & 63.3$\pm$27.6          & 85.6$\pm$06.0          & 55.7$\pm$14.2          & 69.8$\pm$25.1          \\ 
w/ WRM                         & 75.9$\pm$13.2          & 70.8$\pm$26.3          & 88.2$\pm$05.9          & 62.1$\pm$13.3          & 77.5$\pm$19.0          \\
w/ WRM+SMD                        & 76.2$\pm$10.9          & 74.9$\pm$24.4          & 89.4$\pm$04.7          & 61.2$\pm$14.0          & 78.8$\pm$17.5          \\ 
3D-WLDM                        & \textbf{81.4$\pm$09.6} & \textbf{75.9$\pm$21.8} & \textbf{89.7$\pm$05.4} & \textbf{71.1$\pm$10.5} & \textbf{80.1$\pm$17.0} \\ 
\specialrule{0.8pt}{0pt}{0pt}
\end{tabular*}
\end{table}

\subsubsection{Semantic segmentation Analysis} 
To assess the downstream utility of the synthesized CT images produced by different ablation variants, we conducted a segmentation study using the publicly available TotalSegmentator tool~\cite{wasserthal2023totalsegmentator}, which automatically delineates major anatomical structures. Segmentation performance was quantified using the Dice Similarity Coefficient (DSC), with results for representative organs and skeletal structures summarized in Table~\ref{tab_seg}.  Ours model demonstrated substantial improvements in segmentation accuracy compared to the Vanilla, with DSC gains of 17.2\% for the spleen, 12.6\% for the kidneys, 4.1\% for the liver, 15.4\% for the stomach, and 10.3\% for vertebrae. These improvements reflect enhanced anatomical fidelity and structural consistency in the synthesized CT volumes, underscoring the clinical relevance of the proposed framework.

To further illustrate the key contributions of each model component, Fig.~\ref{fig_sig} presents qualitative comparisons of segmented anatomical structures across model variants. The introduction of the SMD module leads to markedly improved spatial alignment of structures such as the femur, reflecting better retention of global anatomical geometry. The addition of WDM and the DSCA module respectively yields enhanced delineation of fine anatomical details, particularly in complex bony regions such as the ribs and vertebrae.  Together, these findings demonstrate that improved synthesis quality directly translates to higher segmentation performance, highlighting the potential of 3D-WLDM to support clinically relevant tasks that depend on structurally accurate CT representations.

\section{CONCLUSION}
We present 3D-WLDM, a novel three-dimensional wavelet latent diffusion model designed for whole-body MR-to-CT modality translation. 
To address the limitations of conventional latent diffusion in cross-modality synthesis, our 3D-WLDM introduces a wavelet-based residual mechanism that enhances the quality of encoding and decoding between image and latent spaces, enabling sharper structural detail and improved tissue contrast. A structure–modality disentanglement strategy further strengthens anatomical consistency by maintaining a persistent structural representation throughout the generative process. Additionally, a dual skip connection attention mechanism refines the fusion of structural and modality-specific features, resulting in greater synthesis stability and reduced artifact prevalence.
Through extensive quantitative and qualitative evaluations, the model demonstrates clear advantages over the state-of-the-art GAN-based and diffusion-based methods, particularly in anatomical fidelity, contrast preservation, and structural consistency.

Beyond performance gains, our 3D-WLDM offers practical benefits for clinical imaging workflows. It enables accurate synthetic CT generation from MR input alone, transforming the inherently challenging MR-to-CT registration task into a mono-modality alignment problem, and supports a range of downstream applications, including PET/MR attenuation correction, MR-only radiation therapy planning, and MR-based organ segmentation using established CT algorithms. The method also provides CT-like visualization capabilities for MR-guided interventions without exposing patients to ionizing radiation. Despite its strengths, the model’s reliance on accurate MR-CT alignment during training may limit robustness in regions affected by anatomical variation or motion. Future work will focus on incorporating semantic- and self-supervised learning strategies to mitigate this sensitivity. Moreover, extending the model’s adaptability to diverse MR sequences and improving computational efficiency and uncertainty quantification will be essential steps toward clinical translation.

\bibliographystyle{IEEEtran}
\bibliography{IEEEabrv,Myreference}

\end{document}